\documentclass{pos}

\newcommand{\bald}[1]{{\bf #1}}

\title{Jet Energy Loss and Mach Cones in pQCD vs. AdS/CFT }
\ShortTitle{Away-side Angular Correlations in pQCD vs. AdS/CFT}

\author{\speaker{Barbara Betz}
         \thanks{We thank S.\ Gubser, A.\ Yarom, and S.\ Pufu for providing the numerical tables for the AdS/CFT
                stress tensor.}\\
        Institut fur Theoretische Physik, Goethe - Universit\"at, 60438 Frankfurt, Germany\\
        E-mail: \email{betz@th.physik.uni-frankfurt.de}}

\author{Jorge Noronha
        \thanks{J.N. and M.G. acknowledge support from DOE under Grant No.
                DE-FG02-93ER40764. }\\
        Department of Physics, Columbia University, New York, 10027, USA\\
        E-mail: \email{noronha@phys.columbia.edu}}
	
\author{Miklos Gyulassy\\
        Department of Physics, Columbia University, New York, 10027, USA\\
        E-mail: \email{gyulassy@phys.columbia.edu}}
	
\author{Giorgio Torrieri\\
        Institut fur Theoretische Physik, Goethe - Universit\"at, 60438 Frankfurt, Germany \\
        E-mail: \email{torrieri@th.physik.uni-frankfurt.de}}	

\abstract{We compare away-side hadron correlations with respect to tagged heavy quark jets
computed within a weakly coupled pQCD and a strongly coupled AdS/CFT model.
While both models feature similar far zone Mach and diffusion wakes,
the far zone stress features are shown to be too weak to survive
thermal broadening at hadron freeze-out. Observable away-side conical correlations
are dominated by the jet-induced transverse flow in near zone ``Neck'' region,
which differs significantly for both models.
Unlike in AdS/CFT, the induced transverse flow in the Neck zone is too weak in pQCD to produce conical correlations
after Cooper-Frye freeze-out. The observation of conical
correlations violating Mach's law would favor the strongly-coupled
AdS/CFT string drag dynamics, while their absence would favor
weakly-coupled pQCD-based hydrodynamics.}

\FullConference{8th Conference Quark Confinement and the Hadron Spectrum\\
		 September 1-6, 2008\\
		 Mainz. Germany}

\begin{document}

\begin{center}
\begin{figure}[th!]
\centerline
\centering
\begin{minipage}[b]{6cm}
\includegraphics[width=6cm]{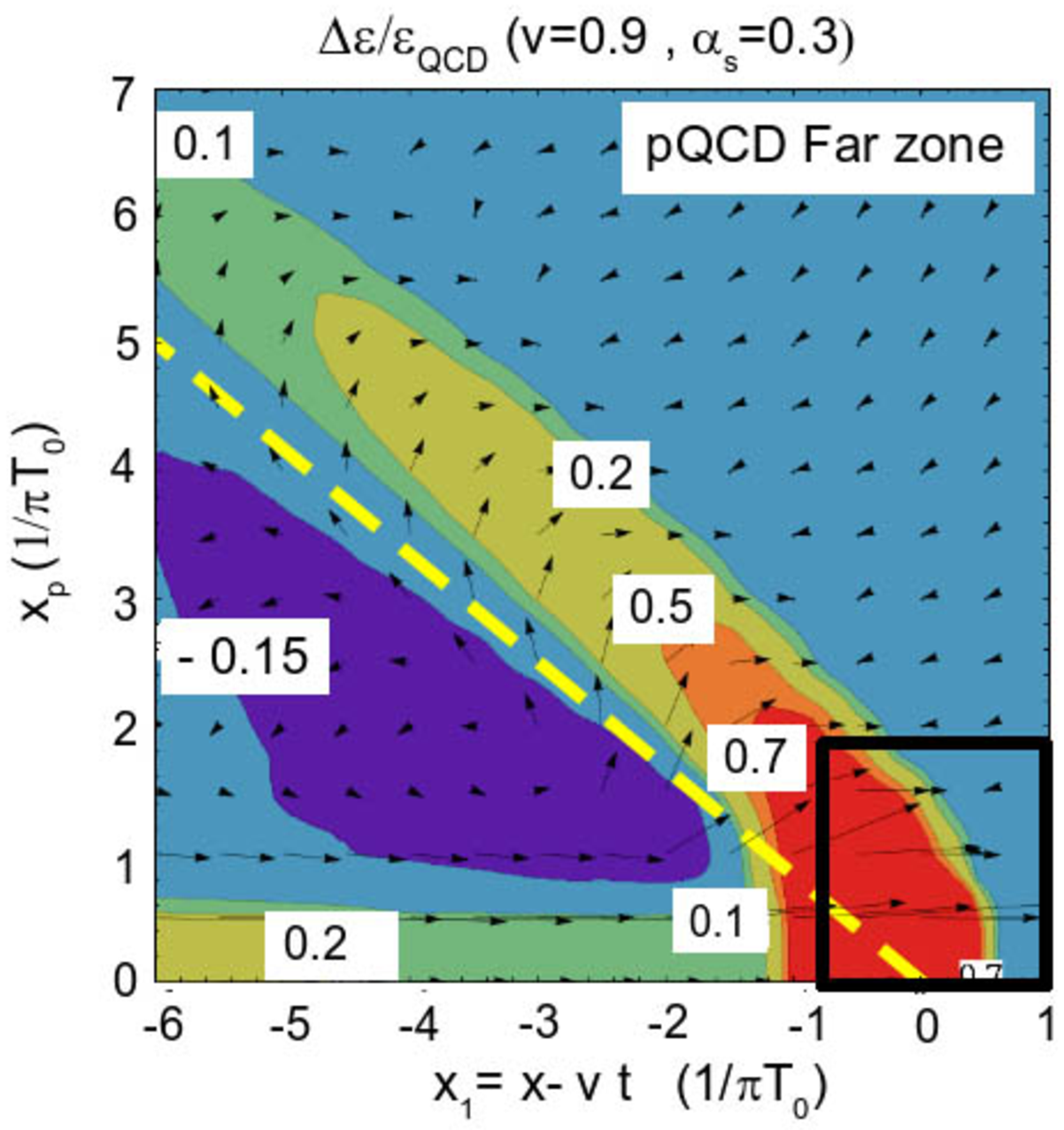}
\end{minipage}
\hspace*{0.5cm}
\begin{minipage}[b]{6cm}
\includegraphics[width=8cm]{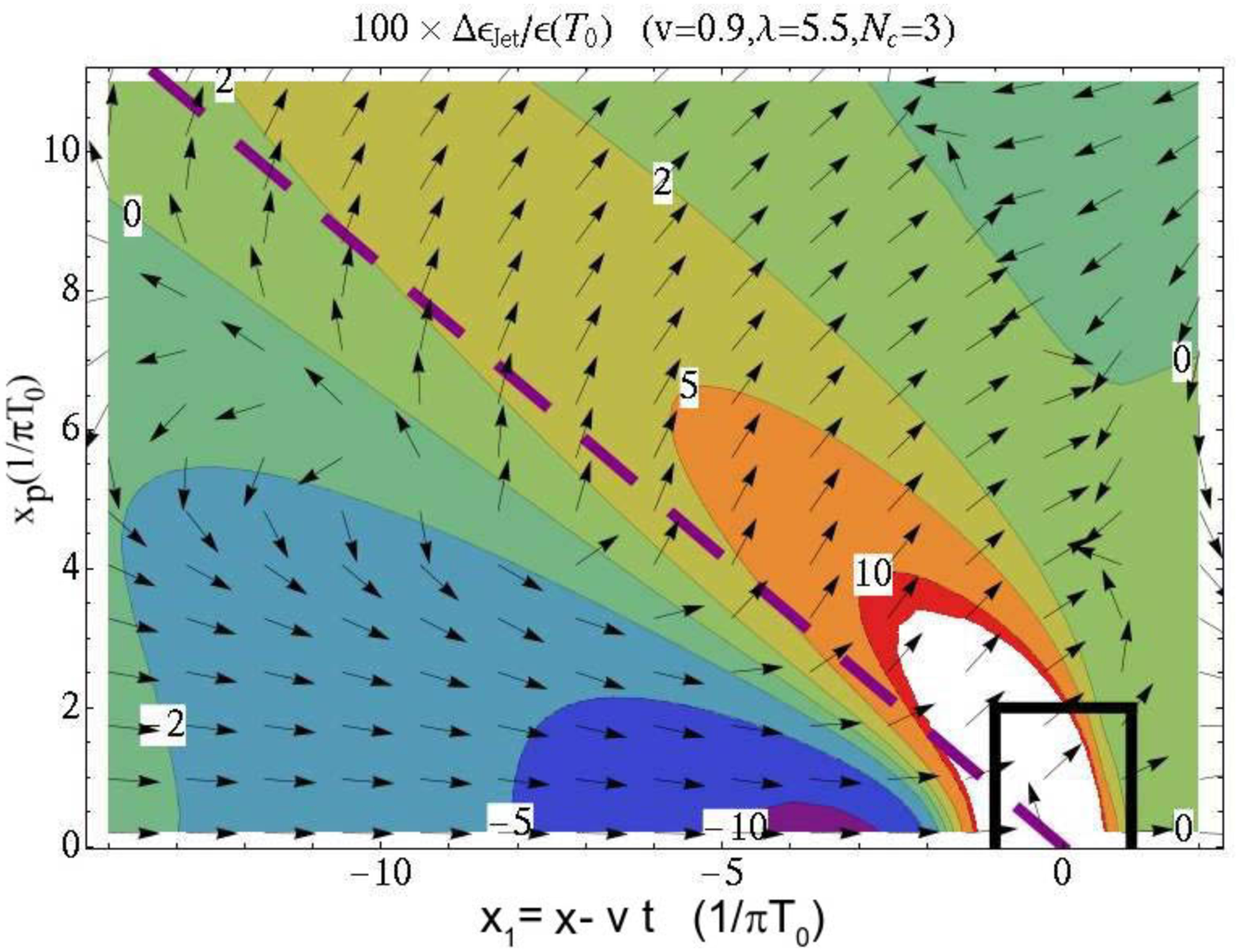}
\end{minipage}
\caption{The fractional energy density perturbation (in the lab frame) due to a heavy quark with $v=0.9$ in a pQCD plasma of temperature
$T_0=200$ MeV, calculated using (3+1)d hydrodynamics with the pQCD source of Neufeld (left panel) and AdS/CFT (right panel).
The ideal Mach cone for a point source is represented by the dashed lines, the box indicates the Neck region.}
\label{farzone}
\end{figure}
\end{center}
\vspace*{-1cm}
\section{Introduction}
Energetic back-to-back jets produced in the early stages of a heavy ion collision traverse the sQGP and deposit energy and momentum along
their path in a way that depends on the non-equilibrium details of the physics of
the field-plasma coupling. In the case when one of the jets is produced near the
surface (the trigger jet), the other supersonic away-side jet moves through the
plasma and generates in the far zone a Mach-like disturbance as seen in
Fig.\ \ref{farzone}. Recent interest in Mach-like di-jet correlations
\cite{Experiment} is due to suggestions \cite{Stoecker:2004qu} that
a measurement of the dependence of the cone angle on the supersonic jet velocity $v$
could provide via Mach's law ($\cos\phi_M= c_s/v$) a constraint on the average speed
of sound in the strongly coupled Quark-Gluon Plasma (sQGP) \cite{Gyulassy:2004zy}
created at the Relativistic Heavy Ion Collider (RHIC). The observation of strong elliptic
flow in non--central Au+Au collisions consistent with fluid dynamical predictions
suggest that a thermalized medium that evolves
hydrodynamically is created in these collisions. Moreover, since the average momentum
of particles emitted on the away--side approaches the value of the thermalized medium
with decreasing impact parameter, the energy lost by the jet should
quickly thermalize \cite{Kolb}. Thus, the disturbance caused by the jet may also be described
hydrodynamically, which requires solving
\begin{eqnarray}
\partial_\mu T^{\mu\nu} =S^\nu
\label{sourceterm}
\end{eqnarray}
to determine the time evolution of the medium that was disturbed by the moving jet. The
source term that correctly depicts the interaction of the jet with the sQGP is unknown from
first principles although recent calculations in pQCD and in AdS/CFT
\cite{AdSCFT} have shed some light on this problem. Here,
we summarize the striking difference between pQCD and AdS/CFT models by solving
numerically the full nonlinear (3+1)d relativistic hydrodynamic equations using the SHASTA hydro
code \cite{Rischke:1995ir}, supplemented with pQCD source derived in Refs.\
\cite{Neufeld,BetzHydro} and comparing the results with those computed in \cite{NGTnonmach}, which used the
$\mathcal{N}=4$ SYM AdS/CFT data computed by Gubser, Pufu, and Yarom in Ref.\ \cite{gubsermach}
for the energy-momentum disturbances caused by the heavy-quark which was created at $t\to -\infty$
and has been moving through the infinitely extended $\mathcal{N}=4$ SYM static background plasma
since then.
For the hydrodynamical results, the initial away-side heavy quark jet is assumed to start $t=0$
at $x_1 =-4.5$ fm and the freeze-out is done (as for the AdS/CFT results) when the heavy quark
reaches the origin of the coordinates at time $t_f=4.5/v$ fm. In the hydro calculations, we specialize to the ideal fluid
case to minimize the dissipative broadening of any conical correlations. Moreover, we neglect here the
near-side associated correlations and use $\alpha_s=1/\pi$ as well as $x_{p\,max}=1/m_D$ as
an infrared cutoff while the minimum lattice spacing naturally provided an ultraviolet cutoff.
The background temperature is set to $T_0=0.2$ GeV.

Given the large theoretical systematic uncertainty inherent in any phenomenological model of hadronization, we consider here two simple limits for modeling the fluid
decoupling and freeze-out. In one often used limit, computational fluid
cells are frozen out via the Cooper--Frye (CF) method where the conversion of the fluid flow velocity field $U^\mu(x)$,
associated (massless) momentum distribution $P^\mu(X)$ and temperature profile $T(X)$ into
free particles (at mid-rapidity) is achieved instantaneously at a critical surface $d \Sigma_\mu$
via
\begin{equation}
\frac{dN}{p_Tdp_Tdy d\phi}\Big
|_{y=0}=\int_{\Sigma}d\Sigma_{\mu}P^{\mu}\left[f_0(U^{\mu},P^{\mu},T)-f_{eq}\right]\, ,
\label{cooperfrye}
\end{equation}
where $p_T$ is the transverse momentum, $\Sigma (X)$ is the isochronous freeze-out hypersurface,
and $f_0=\exp(-U^{\mu}P_{\mu}/T(X))$ is a local Boltzmann equilibrium
distribution. As an independent calorimetric-like observable of collective flow we also
investigate the bulk momentum weighted polar angle distribution (in the laboratory
frame)
\begin{equation}
\frac{d S}{d\cos\theta} = \sum_{cells} |\vec{\mathcal{P}}_c|
\delta\left(\cos\theta- \cos\theta_c\right)\nonumber
= \int d^3 {\bf x}\,\, |\bald{M}(X)|
\delta\left(\cos\theta- \frac{M_x(X)}{|{\bf M}(X)|}\right)\Big|_{t_f}
\label{bulkeq}
\end{equation}
where $\theta=0$ corresponds to the jet direction and $\theta \in [0,\pi]$.
This quantity measures the angular distribution of fluid momentum  at freeze-out time, $t_f$,
in each fluid cell \cite{BetzHydro}. The strong assumption in this decoupling scheme is that the frozen cells
do not break up (unlike in the CF case) but preserve their momentum
$\mathcal{P}^i_c=d^3{\bf x}\,\, T^{0i}({\bf x},t_f)$ to the detector. Here $\theta$ is the polar
angle with respect to the away-side heavy quark jet.

\begin{center}
\begin{figure}[th!]
\vspace*{-0.8cm}
\begin{minipage}[b]{6cm}
\includegraphics[width=6cm]{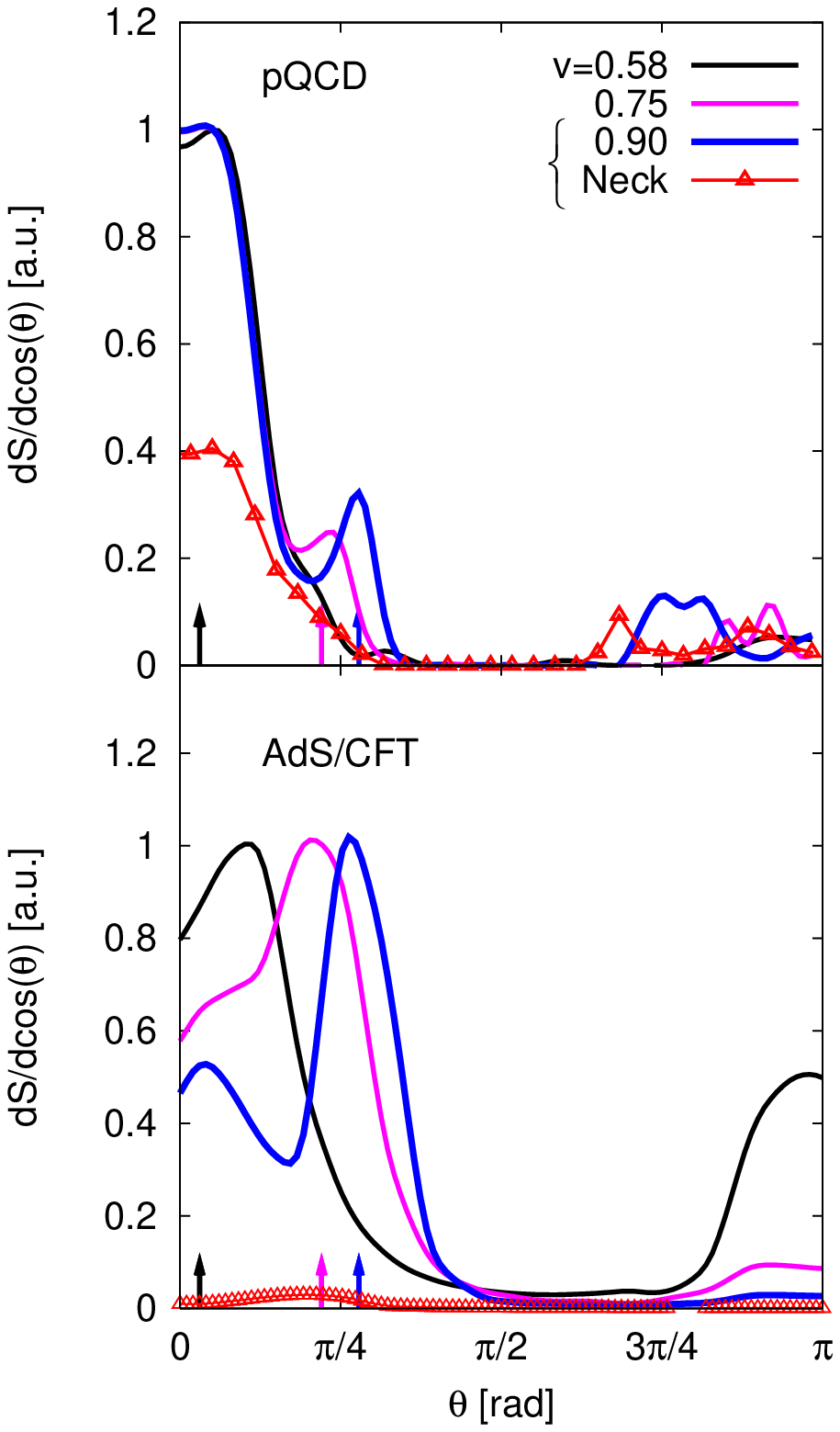}
\end{minipage}
\hspace*{1.5cm}
\begin{minipage}[b]{6cm}
\includegraphics[width=6cm]{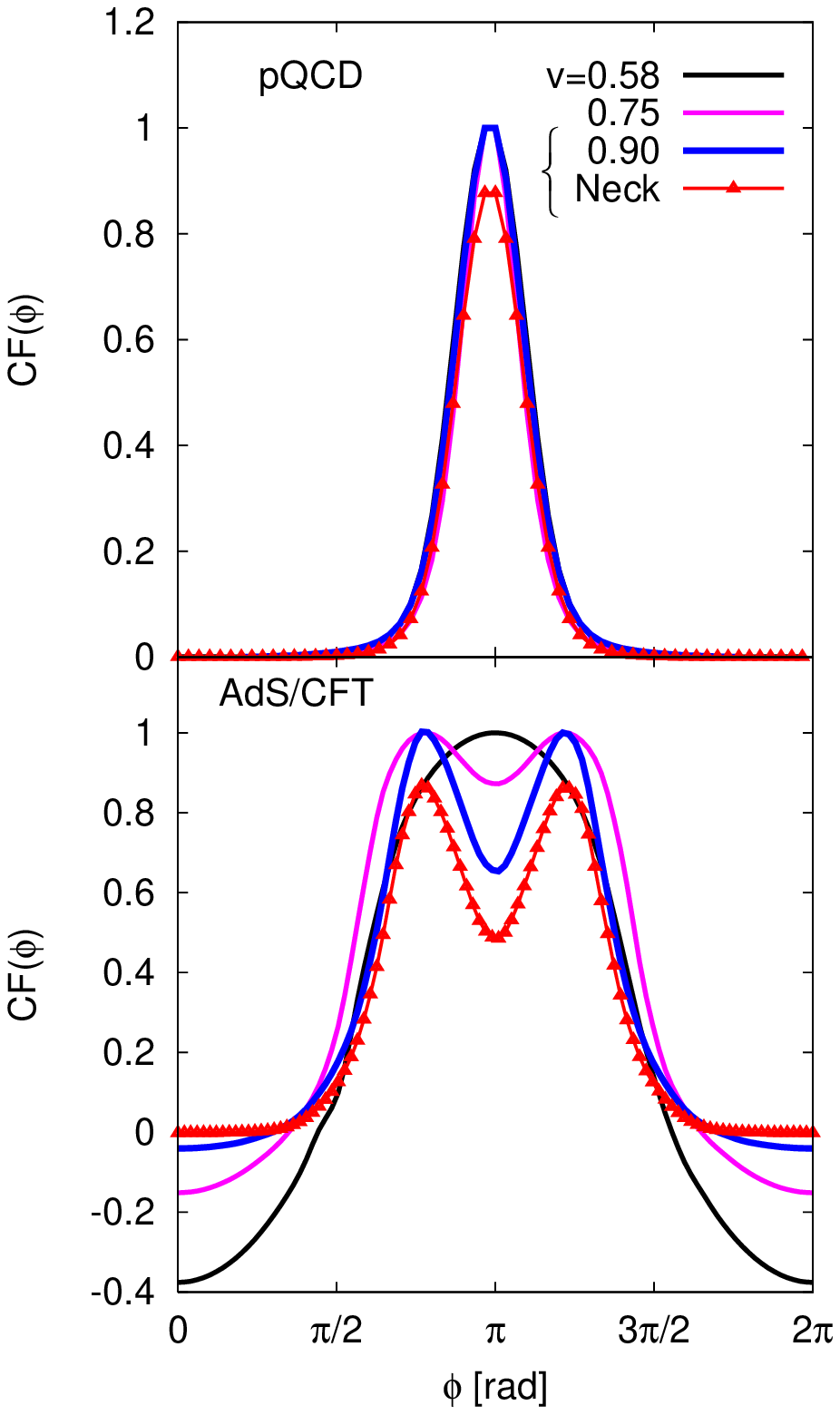}
\end{minipage}
\caption{ (Color online) The (normalized) momentum weighted bulk flow angular distribution (left panel) and
Cooper-Frye freezeout (right panel) for a jet with $v=0.58$ (black), $v=0.75$ (magenta),
and $v=0.90$ (blue) comparing a pQCD and AdS/CFT string drag model. The red line with
triangles represents the Neck contribution for a jet with $v=0.9$ and the arrows indicate the location of the
ideal Mach-cone for $c_s=1/\sqrt{3}$. The negative yield in the lower right panel is due to the presence of
the vortices discussed in the text.}
\label{freezeout}
\end{figure}
\vspace*{-0.5cm}
\end{center}

\section{Comparing pQCD and AdS/CFT}
The results for the normalized bulk flow according to Eq.\ (\ref{bulkeq}) in pQCD
are shown in the upper left panel in Fig.\ \ref{freezeout}.
For all velocities studied here, the pQCD bulk energy flow distribution has a large
forward moving component in the direction of the jet. In the far zone, this forward moving energy
flow corresponds to the diffusion wake. The red curve with triangles in the upper panel in
Fig.\ \ref{freezeout} shows the yield solely from the Neck
region for $v=0.9$, a novel nonequilibrium near zone featured by the AdS/CFT string drag solution
where especially strong transverse flow relative to the jet axis appears (cf.\ box region in Fig.\
\ref{farzone}). The relatively small transverse energy flow in the Neck region is evident
on the left panel of see Fig. \ref{farzone} in contrast to the much larger
transverse flow predicted via AdS in that near zone (right panel of Fig.\ \ref{farzone}).  The Mach cone
emphasized in Ref.\ \cite{Neufeld} is
also clearly seen but its amplitude relative to the mostly forward
diffusion plus Neck contribution is much smaller than in the AdS/CFT
case.
However, when $v=0.58$ the
finite angle from the Mach cone is overwhelmed by the strong bow shock
formed in front of the quark, which itself leads to small conical dip
not at the ideal Mach angle (black arrow).
The bottom left panel in Fig.\ \ref{freezeout}
shows that in the AdS/CFT case more cells are pointing in a
direction near the Mach cone angle than in the forward direction
(diffusion wake) when $v=0.9$ and $v=0.75$. 
The red line with triangles in the
bottom panel of Fig.\ \ref{freezeout} shows that the relative
magnitude of the contribution from the Neck region to the final bulk
flow result in AdS/CFT is much smaller than in pQCD. However, note
that small amplitude peak in the AdS/CFT Neck curve is located at a
much larger angle than the corresponding peak in the pQCD Neck, as one
would expect from the transverse flow shown Fig.\ \ref{farzone}.
Moreover, for all velocities studied here, a peak occurs in
direction of the trigger particle, representing the backward flow that is always
present vortex-like structures created by the jet as
discussed in detail in Ref.\ \cite{Betz:2007kg}.

The right panel of Fig.\ \ref{freezeout} shows our normalized CF freeze-out
results for the associated away-side azimuthal distribution
for light hadrons with $v=0.58,0.75,0.9$ at  mid-rapidity and $p_T=5 \pi \,T_0 \sim 3.14$ GeV.
The pQCD angular distribution displays only a sharp peak at $\phi=\pi$ for all velocities.
Note that the different peaks found in the bulk
flow analysis of the pQCD data shown in the upper left panel in
Fig. \ref{freezeout} do not survive CF freeze-out. We conclude that the strong forward moving diffusion wake
as well as the mostly forward bow shock Neck zone dominate
the away-side peak and that the thermal broadened Mach correlations are too weak in pQCD to contribute to the final angular correlations.
In the AdS/CFT case (see lower right panel of Fig.\ \ref{freezeout}),
a double peak structure can be seen for
$v=0.9$ and $v=0.75$. Note, however, that the peaks in the AdS/CFT
correlation functions do not obey Mach's law. This is because these
correlations come from the Neck region where there is a strong
transversal non-Mach flow \cite{NGTnonmach}. This is explicitly shown
in the red curve with triangles that represents the Neck contribution
for a jet with $v=0.9$ as in Fig.\ \ref{freezeout}. For $v=0.58$,
the resulting flow is not strong enough to lead to non-trivial angular
correlations.

\section{Conclusions}
In this paper, we compared the away-side angular hadron correlations associated with tagged heavy quark jets obtained in pQCD and
AdS/CFT. In both cases the true Mach wakes are not observable after
the standard CF method. Mach-like peaks are only observable in the sudden shattering freeze-out
scenario described in Eqn.\ (\ref{bulkeq}). Moreover, the Neck region gives the largest contribution to
the total yield in CF freeze-out while its contribution in the other extreme case involving the bulk
flow hadronization is not as relevant. We propose that the measurement of the jet velocity
dependence of the associated away-side correlations with identified heavy quark triggers at RHIC and
LHC will provide important constraints on possible on the jet-medium coupling dynamics.

\end{document}